\documentclass[twocolumn,showpacs,preprintnumbers,amsmath,amssymb,prl,superscriptaddress]{revtex4}
\usepackage{graphicx}
\usepackage{dcolumn}
\usepackage{bm}

\begin{document}

\title{Multiplicity and Pseudorapidity Distributions of  Photons in Au + Au Collisions at $\sqrt{s_{NN}}$ = 62.4 GeV}

\medskip

 \affiliation{Argonne National Laboratory, Argonne, Illinois 60439}
 \affiliation{University of Bern, 3012 Bern, Switzerland}
 \affiliation{University of Birmingham, Birmingham, United Kingdom}
 \affiliation{Brookhaven National Laboratory, Upton, New York 11973}
 \affiliation{California Institute of Technology, Pasadena, California 91125}
 \affiliation{University of California, Berkeley, California 94720}
 \affiliation{University of California, Davis, California 95616}
 \affiliation{University of California, Los Angeles, California 90095}
 \affiliation{Carnegie Mellon University, Pittsburgh, Pennsylvania 15213}
 \affiliation{Creighton University, Omaha, Nebraska 68178}
 \affiliation{Nuclear Physics Institute AS CR, 250 68 \v{R}e\v{z}/Prague, Czech Republic}
 \affiliation{Laboratory for High Energy (JINR), Dubna, Russia}
 \affiliation{Particle Physics Laboratory (JINR), Dubna, Russia}
 \affiliation{University of Frankfurt, Frankfurt, Germany}
 \affiliation{Institute  of Physics, Bhubaneswar 751005, India}
 \affiliation{Indian Institute of Technology, Mumbai, India}
 \affiliation{Indiana University, Bloomington, Indiana 47408}
 \affiliation{Institut de Recherches Subatomiques, Strasbourg, France}
 \affiliation{University of Jammu, Jammu 180001, India}
 \affiliation{Kent State University, Kent, Ohio 44242}
 \affiliation{Lawrence Berkeley National Laboratory, Berkeley, California 94720}
 \affiliation{Massachusetts Institute of Technology, Cambridge, MA 02139-4307}
 \affiliation{Max-Planck-Institut f\"ur Physik, Munich, Germany}
 \affiliation{Michigan State University, East Lansing, Michigan 48824}
 \affiliation{Moscow Engineering Physics Institute, Moscow Russia}
 \affiliation{City College of New York, New York City, New York 10031}
 \affiliation{NIKHEF, Amsterdam, The Netherlands}
 \affiliation{Ohio State University, Columbus, Ohio 43210}
 \affiliation{Panjab University, Chandigarh 160014, India}
 \affiliation{Pennsylvania State University, University Park, Pennsylvania 16802}
 \affiliation{Institute of High Energy Physics, Protvino, Russia}
 \affiliation{Purdue University, West Lafayette, Indiana 47907}
 \affiliation{University of Rajasthan, Jaipur 302004, India}
 \affiliation{Rice University, Houston, Texas 77251}
 \affiliation{Universidade de Sao Paulo, Sao Paulo, Brazil}
 \affiliation{University of Science \& Technology of China, Anhui 230027, China}
 \affiliation{Shanghai Institute of Applied Physics, Shanghai 201800, China}
 \affiliation{SUBATECH, Nantes, France}
 \affiliation{Texas A\&M University, College Station, Texas 77843}
 \affiliation{University of Texas, Austin, Texas 78712}
 \affiliation{Tsinghua University, Beijing 100084, China}
 \affiliation{Valparaiso University, Valparaiso, Indiana 46383}
 \affiliation{Variable Energy Cyclotron Centre, Kolkata 700064, India}
 \affiliation{Warsaw University of Technology, Warsaw, Poland}
 \affiliation{University of Washington, Seattle, Washington 98195}
 \affiliation{Wayne State University, Detroit, Michigan 48201}
 \affiliation{Institute of Particle Physics, CCNU (HZNU), Wuhan 430079, China}
 \affiliation{Yale University, New Haven, Connecticut 06520}
 \affiliation{University of Zagreb, Zagreb, HR-10002, Croatia}

 \author{J.~Adams}\affiliation{University of Birmingham, Birmingham, United Kingdom}
 \author{M.M.~Aggarwal}\affiliation{Panjab University, Chandigarh 160014, India}
 \author{Z.~Ahammed}\affiliation{Variable Energy Cyclotron Centre, Kolkata 700064, India}
 \author{J.~Amonett}\affiliation{Kent State University, Kent, Ohio 44242}
 \author{B.D.~Anderson}\affiliation{Kent State University, Kent, Ohio 44242}
 \author{D.~Arkhipkin}\affiliation{Particle Physics Laboratory (JINR), Dubna, Russia}
 \author{G.S.~Averichev}\affiliation{Laboratory for High Energy (JINR), Dubna, Russia}
 \author{S.K.~Badyal}\affiliation{University of Jammu, Jammu 180001, India}
 \author{Y.~Bai}\affiliation{NIKHEF, Amsterdam, The Netherlands}
 \author{J.~Balewski}\affiliation{Indiana University, Bloomington, Indiana 47408}
% ...

 \collaboration{STAR Collaboration}\noaffiliation

\date{\today}

\begin{abstract}
We present the first measurement of multiplicity and pseudorapidity
distribution of photons in the pseudorapidity region 
2.3 $\le$ $\eta$  $\le$ 3.7 for different centralities in Au + Au 
collisions at $\sqrt{s_{NN}}$ = 62.4 GeV. 
We find that the photon yield in this
pseudorapidity range scales with the number of participating nucleons at
all collision centralities studied.
The pseudorapidity distribution of photons, dominated by
neutral pion decays, has been compared to those of identified 
charged pions, photons, and inclusive charged particles from heavy 
ion and nucleon-nucleon collisions at various energies. The photon 
production in the measured pseudorapidity region has been shown 
to be consistent with the energy and centrality independent 
limiting fragmentation scenario.
\end{abstract}

\pacs{25.75.Dw}
\maketitle

 One of the primary goals of the heavy-ion program at the Relativistic
 Heavy-Ion Collider (RHIC) at Brookhaven National Laboratory is to
 search for the possible formation of Quark-Gluon Plasma~\cite{qm01}.
 Important information about the dynamics of particle production and
 the evolution of the system formed in the collision can be obtained from
 various global observables, such as the multiplicity of photons
 and charged particles.  At RHIC energies, the particle production
 mechanisms could be different in different regions of pseudorapidity
 ($\eta$)~\cite{phobos,brahms}.  At midrapidity 
 a significant increase in charged particle production
 normalized to the number of participating nucleons ($N_{\mathrm part}$) 
 has been observed for central Au+Au collisions compared to 
 peripheral Au+Au and p+p collisions~\cite{phenix}.  This has been 
 attributed to the onset of hard  scattering processes, which scale 
 with the number of binary collisions. Alternatively, in the Color 
 Glass Condensate~\cite{cgc} picture of particle production at 
 midrapidity, the centrality dependence could reflect increasing 
 gluon density due to the decrease in the effective strong coupling constant. 
 However, the total charged particle multiplicity per participant pair,
 integrated over the whole pseudorapidity range, is found to be
 independent of centrality in Au+Au collisions ~\cite{phobos}. 

  It is also observed that the number of charged particles produced 
  per participant pair as a function of $\eta$ - y$_{\mathrm beam}$, 
  where y$_{\mathrm beam}$ is the beam rapidity, is independent of 
  beam energy~\cite{phobos}. This phenomenon is known as limiting
  fragmentation (LF)~\cite{limiting_frag}.  There have been
  contradictory results reported from inclusive charged particle
  measurements regarding the centrality dependence of the LF
  behavior, results from PHOBOS show a centrality dependence~\cite{phobos},
  while those from BRAHMS show a centrality independent behaviour
  ~\cite{brahms}.
  The centrality dependence at forward rapidities 
  has been attributed to nuclear remnants and baryon stopping.  
  The role of a new mechanism of baryon production~\cite{baryon_junction} 
  also needs to be understood.  Further insight into this question 
  can be obtained by
  studying the centrality, beam energy and system size dependence of 
  LF phenomena with identified particles. Beam energy independence of LF
  for identified pions has been found in
  $e^{+}$$e^{-}$ collisions~\cite{particledatabook} .

 Photons are produced in all stages of the system created in heavy ion
 collisions. They do not interact strongly with the medium and carry
 information about the history of the collision. 
 Since inclusive photon production is dominated 
 by photons from the decay of $\pi^{0}$'s,
 measurement of the multiplicity of photons 
 is complementary to the charged pion measurements.
 The forward rapidity region in heavy ion collisions, where the present
 measurements have been carried out, constitutes an environment that
 precludes the use of a calorimeter due to the high level of overlap
 of fully developed showers. The only measurements of photon 
 multiplicity distribution in the forward rapidity region 
 reported to date are from a preshower detector~\cite{wa93_nim}
 at the SPS, resulting in the study of various aspects of the 
 reaction mechanism in heavy ion  
 collisions~\cite{wa98_dndy,wa93_dndy}.

In this Letter we present the first measurement of photon production
at the forward rapidities (2.3 $\le$ $\eta$  $\le$ 3.7),
carried out by the STAR experiment using a highly granular preshower 
photon multiplicity detector (PMD)~\cite{starpmd_nim} in Au + Au 
collisions at $\sqrt{s_{NN}}$ = 62.4 GeV.
The STAR experiment consists of several detectors to measure 
 hadronic and electromagnetic observables~\cite{star_nim}. 
The minimum bias trigger is obtained using the charged particle hits
from an array of scintillator slats arranged in a barrel 
called the Central Trigger Barrel surrounding the 
Time Projection Chamber (TPC) and two zero degree hadronic 
calorimeters at $\pm$ 18 m from the detector center~\cite{trigger}.
A total of 334000 minimum bias events, corresponding to 0 to 80\% 
of the Au+Au hadronic interaction  cross section, have been selected 
with a collision vertex position of 
less than 30 cm from the center of the TPC along the beam axis.
The centrality determination in this analysis uses the multiplicity of 
charged particles in the pseudorapidity region 
$\mid\eta\mid$ $<$ 0.5, as measured by the TPC~\cite{star_glauber}.

The PMD is located 5.4 meters away from the center of the TPC 
(the nominal collision point) along the beam axis.
It consists of two planes (charged particle veto and preshower)
of an array of cellular gas proportional counters~\cite{starpmd_nim}. 
A lead plate of 3 radiation length thickness was placed between
the two planes and was used as a photon converter.
The sensitive medium is a gas mixture of Ar and
CO$_2$ in the ratio of 70\%:30\% by weight.
There are 41472 cells in each plane, placed inside 12 high voltage 
insulated and gas-tight chambers called super modules (SM).
A photon traversing the converter produces an electromagnetic shower 
in the preshower plane, leading to a larger signal spread over several 
cells as compared to a charged particle which is essentially confined 
to one cell~\cite{starpmd_nim}. In the present analysis, only the data 
from the preshower plane have been used.

The cell-wise response is obtained by using the ADC distributions of
isolated cells.
The ADC distribution of an isolated cell may be treated as the response
of the cell to charged particles~\cite{starpmd_nim}.
For most of the cells this response followed a Landau distribution. 
We used the mean of the ADC distribution of isolated cells
to estimate and correct the relative gains of all cells within each
SM. The cell-to-cell gain variation within a SM varied between 10 - 25\%
for different SMs.

The extraction of photon multiplicity proceeds in two steps involving  
clustering of hits and photon-hadron discrimination.
Hit clusters consist of contiguous cells.
Photons are separated from charged particles using the
following conditions : (a) The number of cells in a cluster is  $>$ 1 and (b) 
the cluster signal is larger than 3 times the average response 
of all isolated cells in a SM.
The choice of the above condition is based on a detailed 
study of the detector response using Monte Carlo simulations.
The number of selected clusters, called
$\gamma -like$  clusters ($N_{\mathrm \gamma -like}$), in different
SMs for the same $\eta$ coverage is used to evaluate the effect of possible
non-uniformity in the response of the detector.

To estimate the number of photons ($N_{\mathrm \gamma}$) from the detected 
$N_{\mathrm \gamma-like}$ clusters we evaluate the photon reconstruction
efficiency ($\epsilon_{\mathrm \gamma}$) and purity ($f_{p}$) of 
the $\gamma-like$ sample defined~\cite{wa98_dndy} 
as $\epsilon_\gamma  =  N^{\gamma,th} _{cls} / N_\gamma$
and $f_p              =  N^{\gamma,th} _{cls} / N_{\gamma-like}$ respectively.
$N_{cls}^{\mathrm \gamma,th}$ is the number of photon clusters 
after applying the photon-hadron discrimination conditions. 
Both $\epsilon_{\gamma}$ 
and $f_{p}$ are obtained from a detailed Monte Carlo simulation using 
the HIJING event generator (version 1.382)~\cite{hijing} 
with default parameter settings and the detector simulation package 
GEANT~\cite{geant}, which incorporates 
the full STAR detector framework for the period this data was taken. 
The lower limit of photon $p_{T}$ acceptance in the PMD is estimated 
to be 20 MeV/c. Both $\epsilon_{\gamma}$ and $f_{p}$ 
vary with pseudorapidity and centrality. This is
due to variations in particle density, upstream conversions  and detector
related effects. The highest occupancy is about 12\% and 
maximum percentage of split cluster is 9\%.
The photon reconstruction efficiency is determined from simulations to
increase from 42\% to 56\% in central collisions
and from 42\% to 70\% in peripheral collisions as $\eta$ increases 
from 2.3 to 3.7. The  purity of the 
photon sample ranges from 55\% to 62\%, and 
from 63\% to 70\%  for central and  peripheral collisions respectively
as we increase $\eta$ within the above range.

The systematic errors on the photon multiplicity ($N_{\mathrm\gamma}$) 
are  due to 
(a) uncertainty in estimates of  $\epsilon_\gamma$ and $f_p$  values, 
    arising from splitting of clusters and the choice of 
    photon-hadron discrimination conditions and  
(b) uncertainty in $N_{\mathrm\gamma}$  arising from the 
non-uniformity of the detector primarily due 
to cell-to-cell gain variation. 
The error in $N_{\mathrm\gamma}$ due to (a) is
estimated from Monte Carlo simulations to be 9.8\% and 7.7 \% in 
central and peripheral collisions respectively.
The error in $N_{\mathrm\gamma}$ due to (b) is
estimated using average gains for normalization and by studying the 
azimuthal dependence of photon density of the detector in a $\eta$ window.
This is found to be 13.5\% for central and 15\% for peripheral collisions.
The total systematic error in $N_{\mathrm \gamma}$ is $\sim$ 17\% 
for both central and peripheral collisions. The errors are obtained by adding 
systematic and statistical errors in quadrature and are shown in all 
the figures. The statistical errors are small and within the symbol sizes.

\begin{figure}
\begin{center}
\vskip -1.0cm
\includegraphics[scale=0.32]{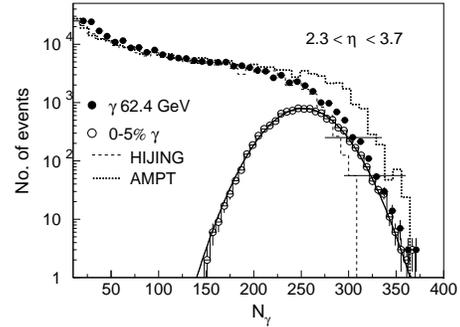}
\caption{Minimum bias $N_{\mathrm \gamma}$ distribution. 
Comparison with HIJING and AMPT models are shown. 
Horizontal bars indicate the errors.
The photon multiplicity distribution for top 5\% central events is shown in open circles.
The solid curve is the fit by a Gaussian.
}
\label{fig1}
\end{center}
\end{figure}


Fig.~\ref{fig1} shows the minimum bias distribution of 
$N_{\mathrm \gamma}$ along with results from HIJING events passed
through detector response (henceforth referred to as HIJING)
and AMPT~\cite{ampt} models. 
The sharp drop in HIJING results at higher $N_{\mathrm \gamma}$ 
is due to lack of statistics.
The HIJING model is based on perturbative QCD processes which lead to 
multiple jet production and jet interactions in matter. 
The AMPT model is a multi-phase transport model which includes both initial
partonic and final hadronic interactions. 
We observe that HIJING underpredicts the measured photon multiplicity whereas
AMPT slightly overpredicts the total measured photon multiplicity for 
central collisions. 
Within the errors, the two models are
in agreement with the measurement.
The top $5\%$ central 
photon multiplicity distribution (open circles) is fitted by a Gaussian with
a mean of 252. 

\begin{figure}
\begin{center}
\vskip -1.1cm
\includegraphics[scale=0.32]{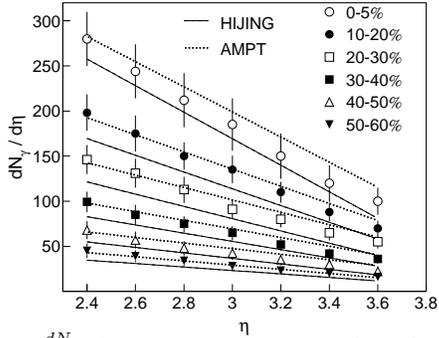}
\vskip -0.5cm
\caption{ $\frac{d N_{\mathrm \gamma}}{d \eta}$ for various event 
centrality classes compared to HIJING and AMPT model calculations.
}
\label{fig2}
\end{center}
\end{figure}

Fig.~\ref{fig2} shows the pseudorapidity distribution of photons
for various event centrality classes. 
The results from HIJING are systematically 
lower compared to data for mid-central and peripheral events. 
The results from AMPT compare well with the data.

\begin{figure}
\begin{center}
\vskip -1.1cm
\includegraphics[scale=0.32]{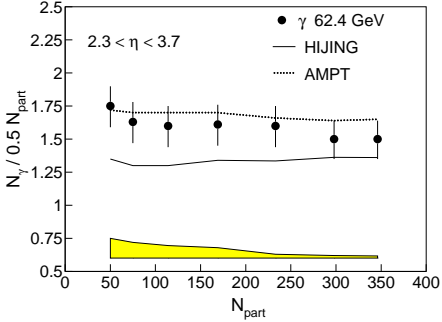}
\vskip -0.3cm
\caption{ Variation of $N_{\mathrm \gamma}$ 
per participant pair in PMD coverage (2.3 $\le$ $\eta$  $\le$ 3.7) as 
a function of $N_{\mathrm part}$. The lower band reflects 
uncertainties in $N_{\mathrm part}$ calculations. 
}
\label{fig3}
\end{center}
\end{figure}

Fig.~\ref{fig3} shows the variation of total number of photons per 
participant pair in the PMD coverage as a function of the number 
of participants. 
$N_{\mathrm part}$ is obtained from Glauber calculations~\cite{star_glauber}. 
Higher values of $N_{\mathrm part}$ corresponds to central collisions.
We observe that the total number of photons per participant
pair is approximately constant with centrality. 
The values from HIJING are lower compared to the data. The values from AMPT 
agree fairly well with those obtained from the data. 
Approximate linear scaling of $N_{\mathrm \gamma}$ 
with $N_{\mathrm part}$ in the $\eta$ range studied 
indicates that photon production is consistent with 
nucleus-nucleus collisions being a superposition of nucleon-nucleon collisions.

\begin{figure}
\begin{center}
\vskip -1.4cm
\includegraphics[scale=0.35]{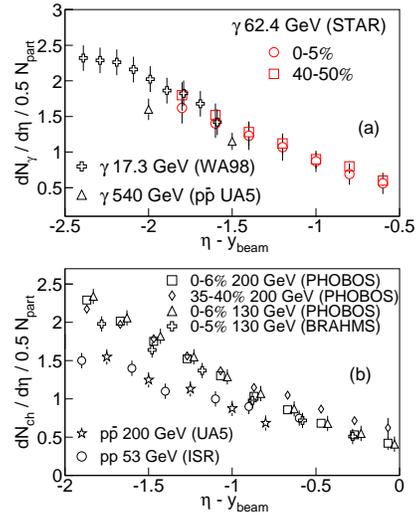}
\vskip -0.3cm
\caption{ (Color Online)
(a) Variation of $\frac{d N_{\mathrm \gamma}}{d \eta}$ normalized
    to  $N_{\mathrm part}$  with 
    $\eta$ - y$_{\mathrm beam}$ 
    for different collision energy and centrality. 
    Also shown $\frac{d N_{\mathrm \gamma}}{d \eta}$ for $p \bar{p}$ 
    collisions.
(b) same as (a) for charged particles.
}
\label{fig4}
\end{center}
\end{figure}
In Fig.~\ref{fig4} we present the energy and centrality dependence of LF
for inclusive photons and charged particles.
Fig.~\ref{fig4}(a) compares the photon pseudorapidity distributions 
for central (0-5\%) and peripheral (40-50\%) Au + Au collisions 
at $\sqrt{s_{NN}}$ = 62.4 GeV, with the top SPS energy central (0-5\%) 
photon data for Pb + Pb collisions~\cite{wa98_dndy} 
as a function of $\eta$ - y$_{\mathrm beam}$. Also shown are the
photon data from  $p \bar{p}$ collisions 
at $\sqrt{s_{NN}}$ = 540 GeV~\cite{ua5}. 
In Fig.~\ref{fig4}(b) we show the charged particle pseudorapidity
distributions for central  (0-6\%),  peripheral (35-40\%) Au + Au collisions 
at $\sqrt{s_{NN}}$ = 200 GeV and  central data at $\sqrt{s_{NN}}$ = 130 GeV
from PHOBOS~\cite{phobos} and BRAHMS~\cite{brahms} as a function of 
$\eta$-y$_{\mathrm beam}$.
Also shown are the
charged particle data from $pp$ and $p \bar{p}$
collisions at $\sqrt{s_{NN}}$ = 53 and 200 GeV~\cite{ua5}. 
We observe in Fig.~\ref{fig4}(a) that photon results from the 
SPS and RHIC are 
consistent with each other, suggesting that photon production follows 
an energy independent LF behavior. Similar energy independent LF behavior
had been observed for charged particles~\cite{brahms,phobos}. 
This can again be seen in Fig.~\ref{fig4}(b) from the comparison 
of charged particle $\eta$ distributions 
from PHOBOS for $\sqrt{s_{NN}}$ = 130 and 200 GeV and 
BRAHMS at $\sqrt{s_{NN}}$ = 130 GeV .

In Fig.~\ref{fig4}(a) we also observe that within the measured $\eta$
range the photon distribution as a function of 
$\eta$ - y$_{\mathrm beam}$ is independent of centrality.
However, in Fig.~\ref{fig4}(b) it is observed that
the charged particle distribution as a function of 
$\eta$ - y$_{\mathrm beam}$ is dependent on centrality~\cite{phobos}.
This centrality dependent behavior of LF observed by PHOBOS 
is most prominent at the lower energy of 
$\sqrt{s_{NN}}$ = 19.6 GeV~\cite{phobos}.
The centrality dependence has been speculated to be due to
nuclear remnants and baryon stopping~\cite{phobos,baryon_junction}. 
The dependence of LF  on the collision system is most clearly seen 
in the comparison between results from heavy ion collisions with 
those from $pp$ and $p \bar{p}$ collisions. We observe in Fig.~\ref{fig4}(a)
that the photon results in the forward rapidity region from $p \bar{p}$ 
collisions at $\sqrt{s_{NN}}$ = 540 GeV are in close agreement with 
the measured photon yield in Au+Au collisions at $\sqrt{s_{NN}}$ = 62.4 GeV.
However the $pp$ and $p \bar{p}$ inclusive charged particle results are 
very different from those reported by PHOBOS (Fig.~\ref{fig4}(b)). 
This indicates that there is apparently a significant charged baryon 
contribution in nucleus-nucleus collisions at forward $\eta$ region.


\begin{figure}
\begin{center}
\vskip -1.4cm
\includegraphics[scale=0.35]{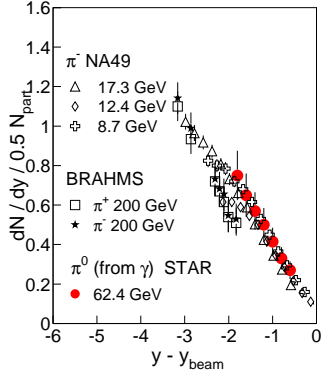}
\vskip -0.3cm
\caption{(Color Online) Estimated $\frac{d N_{\mathrm \pi^{0}}}{dy}$
from $\frac{d N_{\mathrm \gamma}}{dy}$ normalized to $N_{\mathrm part}$,
as compared to $\frac{d N_{\mathrm \pi^\pm}}{dy}$ normalized to  $N_{\mathrm part}$, as a function of y - y$_{\mathrm beam}$ for central collisions at various collision energies.
}
\label{fig5}
\end{center}
\end{figure}
 Fig.~\ref{fig5} shows the charged pion rapidity density in Au+Au collisions
 RHIC~\cite{rhic_pion} and  Pb+Pb collisions at the SPS~\cite{na49_pion}
 and estimated  $\pi^{0}$  rapidity density from the present 
 measurement (photon rapidity density)
 at $\sqrt{s_{NN}}$ = 62.4 GeV, all as a function of y-y$_{\mathrm beam}$. 
 HIJING calculations indicate that about 93-96\% of photons are from $\pi^0$
 decays.  From HIJING we obtained the ratio of the photon to $\pi^0$ yields.
 This ratio is used to estimate the $\pi^0$ yield from the 
 measured photon yield.
 The BRAHMS results at forward rapidities are slightly lower compared to
 the results from SPS energies. However, in general, the results show
 that pion production in heavy ion collisions in the fragmentation
 region agrees with the LF picture. Similar features have been observed
 in $e^{+}$$e^{-}$ collisions~\cite{particledatabook}. The centrality
 dependence of LF for inclusive charged hadrons and the centrality
 independence of limiting fragmentation for identified mesons
 indicate that although the baryon stopping is different in different
 collision systems, the pions produced at forward rapidities
 are not affected by the baryon transport. 

In summary, we have presented the first results of photon multiplicity
measurements at RHIC in the pseudorapidity region 2.3 $\le$ $\eta$  $\le$ 3.7. 
The pseudorapidity distributions of photons
have been obtained for various centrality classes. 
Photon production per participant pair is 
found to be approximately independent of centrality in this 
pseudorapidity region. Comparison with photon and charged 
pion data at RHIC and SPS energies shows, for the
first time in heavy ion collisions, that photons and pions 
follow an energy independent limiting fragmentation behavior, as 
previously found for inclusive charged particles. 
Furthermore, photons are observed to follow a centrality 
independent limiting fragmentation scenario.

\begin{acknowledgments}        
We thank the RHIC Operations Group and RCF at BNL, and the
NERSC Center at LBNL for their support. This work was supported
in part by the HENP Divisions of the Office of Science of the U.S.
DOE; the U.S. NSF; the BMBF of Germany; IN2P3, RA, RPL, and
EMN of France; EPSRC of the United Kingdom; FAPESP of Brazil;
the Russian Ministry of Science and Technology; the Ministry of
Education and the NNSFC of China; SFOM of the Czech Republic,
FOM and UU of the Netherlands, DAE, DST, and CSIR of the Government 
of India; the Swiss NSF; the Polish State Committee for Scientific 
Research; STAA of Slovakia, and the Korea Sci. \& Eng. Foundation.
We acknowledge the help of CERN for use of GASSIPLEX chips in the PMD readout.
\end{acknowledgments}

\normalsize


\end{document}